\documentclass{article}%
\usepackage{amsfonts,amssymb}%
\usepackage{amsmath}%
\usepackage{amsfonts}%
\usepackage{amssymb}%
\usepackage{graphicx}
\newtheorem{defin}{\bf Definition}[section]
\newtheorem{thm}[defin] {\bf Theorem}

\newtheorem{ex}[defin]  {\bf Example}
\newenvironment{proof}{\noindent\mbox{\sc Proof: }}{\rm\hspace*{\fill}$\Box$\vspace{10pt}}
\begin{document}

\title{\textbf{Extending states on finite concrete logics}}
\author{\textbf{Anna De Simone}, \textbf{Mirko Navara} and \textbf{Pavel Pt\'{a}k}
\thanks{The first author acknowledges the support by the European Union under
project Miracle ICA 1- CT-2000-70002 and GNAMPA of INdAM. The second author is
grateful to the grant 201/03/0455 of the Grant Agency of the Czech Republic.
The third author acknowledges the support of the Czech Ministry of Education
under Research Programme MSM 212300013 ``Decision Making and Control in
Manufacturing". }}
\date{}
\maketitle

\begin{abstract}
In this note we collect several observations on state extensions. They may be
instrumental to anyone who pursues the theory of quantum logics. In
particular, we find out when extensions (resp. signed extensions) exist in the
``concrete" concrete logic of all even-element subsets of an even-element set
(Th~\ref{ext-even} and Th~\ref{ext-sub}). We also mildly add to the study of
difference-closed logics as initiated in \cite{ovc} by finding an extension
theorem for subadditive states. Our results suplement the research carried on
in \cite{ads}, \cite{gu}, \cite{gm}, \cite{napt}, \cite{na1}, \cite{ovc},
\cite{pt1}, \cite{pp}, \cite{she}, \cite{sul}, and \cite{svo}.

\vspace{1em} \noindent\textit{2000 AMS Classification:\/} 06C15,
81P10.\newline\textit{Key words:\/} (concrete) quantum logic, state, extension.

\end{abstract}

\section{Introduction}

The question of extending states on quantum logics is sometimes surprisingly
combinatorially involved. In spite of the progress made by the authors
referred to in the abstract above, several questions remain open (see e.g.
\cite{ovc} and \cite{pt1}). It therefore seems helpful to have the situation
clarified in the ``testing" case of the logic of even-number-element subsets
of a set. This is what we intend to do in this note. Our results may partially
overlap with the results of the previous effort but we are not aware of them
being explicitly formulated elsewhere.


\section{Notions and results}

We shall exclusively deal with finite concrete (= set representable) quantum
logics and states (= probability measures) on them as defined below.
Standardly, for a set $X$ we stand $\exp(X)$ for the (Boolean) power algebra
of $X$.

\begin{defin}
\label{cql} Let $X$ be a finite set. A collection $\Delta\subseteq\exp X$ is
said to be a \emph{concrete quantum logic} (abbr. a \emph{logic}) if the
following conditions are satisfied:

\begin{enumerate}
\item $X\in\Delta$;

\item if $A\in\Delta$, then $A^{c}\in\Delta$ ($A^{c}$ denotes the complement
of $A$ in $X$);

\item if $A,B\in\Delta$ and $A\cap B=\emptyset$, then $A\cup B\in\Delta$.
\end{enumerate}

Let $m:\Delta\rightarrow I\!\!R$ be a mapping from a logic into the set of
real numbers. We say that $m$ is a \emph{signed measure} on $\Delta$ if
$m(A\cup B)=m(A)+m(B)$ provided that $A,B\in\Delta$ and $A\cap B=\emptyset$. A
signed measure $m:\Delta\rightarrow I\!\!R$ is said to be a \emph{state} if
$m(A)\geq0$ for any $A\in\Delta$ and $m(X)=1$.
\end{defin}

The chief question we ask here reads as follows: Having given a (concrete)
logic $\Delta$ on a set $X$ and having given a state $s$ on $\Delta$, when can
we extend $s$ as a state (resp. as a signed measure) over the entire algebra
$\exp X$? Expressed more formally, given a state $s$ on $\Delta$, when can we
find a state $t$ (resp., a signed measure $t$) on the Boolean algebra $\exp X$
such that $t$ restricted to $\Delta$ equals to $s$? Let us observe first that
certainly not always. In fact, we even do not have the ``weak" extensions (=
extensions of states to signed measures) of two-valued states at our disposal
as the following simple example shows.

\begin{ex}
\label{no-ext} Let $X=\{1,2,\ldots,6\}$, and let $\Delta$ be the smallest
concrete logic on $X$ containing the following four sets:%
\[
A=\{1,2,3\},\qquad B=\{2,3,4\},\qquad C=\{3,4,5\},\qquad D=\{1,3,5\}.
\]
Obviously, $\Delta=\{\emptyset,A,A^{c},B,B^{c},C,C^{c},D,D^{c},X\}$ (for the
reader acquainted with the theory of orthomodular lattices, $\Delta$ is a
representation of the orthomodular lattice $MO_{4}$). Let us take the
(two-valued) state $s:\Delta\rightarrow\{0,1\}$ defined as follows:
\[
s(A)=0,\quad s(B)=1,\quad s(C)=0,\quad s(D)=1.
\]
We claim that this state $s$ cannot be extended over $\exp X$ as a signed
measure. Indeed, suppose that $m:\Delta\rightarrow I\!\!R$ is a signed measure
which extends $s$. Then%
\[
m(A)+m(B^{c})+m(C)+m(D^{c})=2m(X),
\]
and, analogously,%
\[
m(A^{c})+m(B)+m(C^{c})+m(D)=2m(X).
\]
But if in the left-hand side we replace $m$ by $s$, we obtain $0$ in the first
equality, and we obtain $4$ in the second. A contradiction, thus, such an
extension $m$ of $s$ does not exist.
\end{ex}

It turns out, however, that many concrete logics do allow for the latter kind
of extension. Such are for instance the logics of the following type (the
conceptual value of these logics within various questions of quantum theories
has been indicated in \cite{gu}, \cite{pp}, \cite{svo}, etc.). Given a finite
set $X$ of even cardinality, we denote by $X_{even}$ the concrete logic of all
subsets of $X$ whose cardinality is even.

\begin{thm}
\label{ext-even} Let $X$ be a finite set of an even cardinality. Let $s$ be a
state on $X_{even}$. Then $s$ can be extended as a signed measure over $\exp
X$.
\end{thm}

\begin{proof}
We can suppose that $\mbox{\rm card}X\geq4$\ (otherwise, the result is
trivial). Take an arbitrary triple $x,u,v$ of distinct elements of $X$. Let
\[
\overline{m}(x)=\frac{1}{2}(s(\{x,u\})+s(\{x,v\})-s(\{u,v\})).
\]
We claim that $\overline{m}(x)$ is independent of the choice of $u,v$.
Further, we claim that letting $x$ vary in $X$, we have defined a mapping
$\overline{m}:X\rightarrow I\!\!R$ with the property that upon setting
$m(A)=\sum_{a\in A}\overline{m}(a)$ for any subset $A$ of $X$, this $m$
constitutes a signed measure which extends $s$.

Let us first check that our definition of $\overline{m}$ is correct, i.e. let
us show that $\overline{m}(x)$ does not depend upon the choice of $u,v$. Take
first a couple $v,w$ such that $x,u,v,w$ are distinct. Let us show that both
the couples $u,v$ and $v,w$ define the same $\overline{m}(x)$. Write
\[
f(x,u,v)=s(\{x,u\})+s(\{x,v\})-s(\{u,v\}).
\]
We want to show that $f(x,u,v)=f(x,v,w)$. Let us compute the difference
\begin{align*}
f(x,u,v)-f(x,v,w)  & =s(\{x,u\})+s(\{x,v\})-s(\{u,v\})\\
& -s(\{x,v\})-s(\{x,w\})+s(\{v,w\})\\
& =\bigl(s(\{x,u\})+s(\{v,w\})\bigr)-\bigl(s(\{x,w\})+s(\{u,v\})\bigr)\\
& =s(\{x,u,v,w\})-s(\{x,u,v,w\})=0.
\end{align*}
Further, for a general element $t\in X$ distinct from $x,u,v,w$, we have
$f(x,u,v)=f(x,v,w)=f(x,w,t)$, proving independence of the value $\overline
{m}(x)$ on the choice of $u$ and $v$. Let us check that $m$ defined as above
extends $s$. Take a set $A=\{x,y\}\subseteq X$. Then we can find $u$ and $v$
such that $x,u,v,w$ are distinct and therefore
\begin{align*}
m(A)  & =\overline{m}(x)+\overline{m}(y)=\frac{1}{2}%
\bigl(f(x,u,v)+f(y,u,v)\bigr)\\
& =\frac{1}{2}%
\bigl(s(\{x,u\})+s(\{x,v\})-s(\{u,v\})+s(\{y,u\})+s(\{y,v\})-s(\{u,v\})\bigr)\\
& =\frac{1}{2}\bigl(2s(\{x,y,u,v\})-2s(\{u,v\})\bigr)=s(\{x,y\}).
\end{align*}
The proof is complete.
\end{proof}

Let us comment shortly on the previous result. Firstly, in the logic
$X_{even}$ we generally cannot extend states as states (see also \cite[Th.
3.5.1(v)]{bha}).

\begin{ex}
\label{tre} Let $X=\{1,2,\ldots,2k\}$, $k\in\mathbb{N}$, $k\geq2$. Consider
the state $s$ on $X_{even}$ such that $s(\{1,c\})=0$ and $s(\{b,c\})=\frac
{1}{k-1}$ for all $b,c\in X\setminus\{1\}$, $b\neq c$. If $m $ is any signed
measure on $\exp X$ extending $s$ we have $m(\{1\})=-\frac{1}{2(k-1)}$ and
$m(\{c\})=\frac{1}{2(k-1)}$ for all $c\in X\setminus\{1\}$.
\end{ex}

Secondly, observing that each two-valued state on $X_{even}$, with
$\mbox{\rm card}\,X\geq6$, has to be a Dirac state (i.e. a state sitting in a
point), we see that there must be extreme states on $X_{even}$ which are not
two-valued (in fact, in \cite{napt} we have constructed some). And, thirdly,
it is worthwhile observing that Th. \ref{ext-even} can also be proved, like
many extension theorems which happen to hold, by the well-known criterion of
Horn and Tarski (see e.g. \cite[Def. 3.2.1, 3.2.2, and Th. 3.2.5, 3.2.10]%
{bha}). In the case of Th.~\ref{ext-even} it would present another proof of a
similar complexity. A minor advantage of the proof method used in
Th.~\ref{ext-even} is that it gives the result for group-valued measures for
the groups which allow for dividing by $2$. Observe in passing that, for
instance, the extension problem of course-grained measures as treated in
\cite{gm} and \cite{ovc} finds in our opinion the Horn-Tarski criterion an
effective proof device (see \cite{dspt}).

The logics of $X_{even}$ present a distinguished example of so-called
difference-closed concrete logics. The latter logics were introduced in
\cite{ovc} under the name of symmetric concrete logics. In this note, let us
use our notation (in our opinion more suggestive).

\begin{defin}
\label{dcl} A concrete logic $\Delta$ on a set $X$ is said to be
\emph{difference-closed }if it is closed under the formation of symmetric
differences, i.e., if for any couple of $A,B\in\Delta$ we have
\[
A\delta B=(A\setminus B)\cup(B\setminus A)\in\Delta\,.
\]

\end{defin}

It seems natural to conjecture that Th \ref{ext-even} can be generalized to
all (finite) difference-closed logics. However, this is not the case.

\begin{ex}
\label{nine} Let $X=\{0,1,2,\ldots,9\}$. Consider the smallest concrete logic
$L$ of subsets of $X$ which is difference-closed and contains the sets
$A=\{0,1,4,7\},\quad B=\{0,2,5,8\},\quad C=\{0,1,2,3\}$ and $D=\{0,4,5,6\}$.
(For the reader acquainted with the theory of orthomodular lattices, the logic
$L$ is a representation of $MO_{15}$(thus, difference-closed logics can be
lattices without being Boolean). The logic $L$ has 15 mutually non-disjoint
elements consisting of 4 points and 15 elements consisting of 6 points that
constitute the complements of the former ones).

Let $s$ be a state on $L$ such that $s(A)=s(C)=s(A\delta B)=1$ and
$s(B)=s(D)=s(C\delta D)=0$ (there obviously are such states). Let us show that
$s$ cannot be extended as a signed state on $\exp X$.

Looking for a contradiction, suppose that $m$ is such an extension. Since
$A\cap B=C\cap D=\{0\}$, we obtain%
\[
2m(\{0\})=s(A)+s(B)-s(A\delta B)=0,
\]
and, also,%
\[
2m(\{0\})=s(C)+s(D)-s(C\delta D)=1.
\]
We have reached a contradiction, verifying the required property.
\end{ex}

When we find ourselves within the area of difference-closed concrete logics,
it seems of interest to deal with rather special states.

\begin{defin}
\label{subadd} Let $L$ be a difference-closed concrete logic on a set $X$ and
let $s:\Delta\rightarrow\lbrack0,1]$ be a state on $\Delta$. We say that $s$
is subadditive if $s(A\delta B)\leq s(A)+s(B)$ for any pair $A,B\in\Delta$.
\end{defin}

The following result indicates the meaning of subadditivity in our context.

\begin{thm}
\label{sub-st} Let $X$ be a finite set and let $L=(X,\Delta)$ be a
difference-closed concrete logic. Let $s$ be a state on $\Delta$ which allows
for an extension, $m$, over $\exp X$ as a signed measure. Let the
intersections of pairs of sets of $L$ generate all atoms of a Boolean algebra
on $X$. Then $s$ is subadditive if and only if $m$ is a state.
\end{thm}

\begin{proof}
Compute $s(A)+s(B)-s(A\delta B)$. We obtain
\begin{align*}
& s(A)+s(B)-s(A\delta B)\\
& =m(A\setminus B)+m(A\cap B)+m(B\setminus A)+m(A\cap B)-m(A\setminus
B)-m(B\setminus A)\\
& =2m(A\cap B)\,.
\end{align*}
We therefore see that $m(A\cap B)$ is nonnegative precisely when
$s(A)+s(B)-s(A\delta B)$ is nonnegative, which occurs precisely when $s$ is
subadditive. We see that the extension $m$ is a state on the Boolean algebra
generated by $\Delta$. By the classical theorem, $m$ can be further extended
over $\exp X$ as a measure. The proof is complete.
\end{proof}

Getting back to Th~\ref{ext-even} interplayed with Th~\ref{sub-st}, we obtain
the following corollary.

\begin{thm}
\label{ext-sub} Let $X$ be a set of even cardinality.
Let $s$ be a state on $X_{even}$. Then $s$ can be extended over $\exp X$ as a
state if and only if $s$ is subadditive.
\end{thm}

A natural question arises if a difference-closed logic allows for extensions
of states over $\exp X$ as signed measures. Regretfully, it is not the case.
We demonstrate it in our final result.

\begin{ex}
\label{sub-no} Let us again consider the logic $L$ of the previous
Ex.~\ref{nine}. Let us consider the state $s$ defined by setting
\[
s(A)=\frac{1}{3},\qquad s(B)=\frac{1}{4},\qquad s(C)=\frac{2}{5}%
,\qquad\hbox{and}
\]%
\[
s(Y)=\frac{1}{2}\quad\hbox{ for any }\quad Y\notin\{A,B,C,A^{c},B^{c},C^{c}\}.
\]
It is not difficult to check that $s$ is a subadditive state on $L$. This
state $s$ cannot be extended over $\exp X$ as a signed measure. Indeed, let
$m$ be such an extension. Then%
\[
2\,m(\{0\})=m(A)+m(B)-m(A\delta B)=s(A)+s(B)-s(A\delta B)=\frac{1}{3}+\frac
{1}{4}-\frac{1}{2}=\frac{1}{12}\,,
\]
whereas, analogously,%
\[
2\,m(\{0\})=m(C)+m(D)-m(C\delta D)=s(C)+s(D)-s(C\delta D)=\frac{2}{5}+\frac
{1}{2}-\frac{1}{2}=\frac{2}{5}\,.
\]
This is a contradiction.
\end{ex}

\bigskip\textbf{Acknowledgement. }The authors want to express their gratitude
to the referee for valuable suggestions which corrected the original version.

\vspace{0.5cm}

\noindent Anna De Simone\newline Dipartimento di Matematica e
Statistica\newline Universit\'a Federico II di Napoli\newline Complesso Monte
S. Angelo\newline Via Cintia - 80126 Napoli Italy\newline email: annades@unina.it

\vspace{0.5cm}

\noindent Mirko Navara\newline Center for Machine Perception\newline
Department of Cybernetics\newline Faculty of Electrical Engineering\newline
Czech Technical University\newline Technick\'{a} 2\newline166 27 Prague, Czech
Republic\newline email: navara@cmp.felk.cvut.cz

\vspace{0.5cm}

\noindent Pavel Pt\'{a}k\newline Department of Mathematics\newline Faculty of
Electrical Engineering\newline Czech Technical University\newline
Technick\'{a} 2\newline166 27 Prague, Czech Republic\newline email: ptak@math.feld.cvut.cz

\end{document}